\documentclass[12pt]{iopart}
\usepackage{bm,amssymb,graphicx}
\usepackage{iopams}

\newcommand{\mn}[1]{{\mathbf #1}}
\newcommand{\hnn}[1]{\widehat{\boldsymbol #1}}

\newcommand{\braket}[2]{\langle #1\vert #2 \rangle}

\begin{document}

\title[Heat flux of a granular gas with homogeneous temperature]{Heat flux of a granular gas with homogeneous temperature}

\author{Nagi Khalil}

\address{IFISC (CSIC-UIB), Instituto de F\'isica Interdisciplinar y Sistemas Complejos, Campus Universitat de les Illes Balears, 
E-07122, Palma de Mallorca, Spain.} \ead{nagi@ifisc.uib-csic.es}

\begin{abstract} 
A steady state of a granular gas with homogeneous granular temperature, no mass flow, and nonzero heat flux is studied. The state is created by applying an external position--dependent force or by enclosing the grains inside a curved two--dimensional silo. At a macroscopic level, the state is identified with one solution to the inelastic Navier--Stokes equations, due to the coupling between the heat flux induced by the density gradient and the external force. On the contrary, at the mesoscopic level, by exactly solving a BGK model or the inelastic Boltzmann equation in an approximate way, a one--parametric family of solutions is found. Molecular dynamics simulations of the system in the quasi--elastic limit are in agreement with the theoretical results.
\end{abstract}



\maketitle

\section{Introduction \label{sec1}}

The rich phenomenology of Granular Matter, usually defined as systems composed of macroscopic particles that undergo inelastic collisions, is frequently exemplified using grains inside containers and silos \cite{janabe96,ka99,arts06}. At mechanical equilibrium, all particles stay at rest and the system resembles a solid. If we open an orifice at the bottom of the silo, grains may start flowing and the system would behave as a fluid. Even more, the system may become a \emph{granular} gas if we shake the silo and grains gain enough energy. The static properties as well as the discharging process of silos has attained much attention in the last years, mostly through experimental and numerical investigations \cite{gogo04,ciconi06,arci14,zupugama03,jazuma12,gucobodapebrcl15}. Surprisingly, little theoretical results for the silo problem can be found in the literature, even at the dilute limit where there is a fundamental theory of Granular Matter based on Kinetic Theory. One of the objectives of the present work is to contribute to the theoretical study of confined grains by considering a special state of a granular gas inside a curved two-dimensional silo. 

Under rapid flow conditions, and using different models of the granular gas, Kinetic Theory of molecular gases has been modified to account for the inelastic collisions \cite{brdukisa98,gadu99,krsaga14}. This mesoscopic description, together with an approximate method of solution to the kinetic equation, provides a hydrodynamic or macroscopic one, that usually involves the number density $n$, the velocity flow $\mn{u}$, and the mean kinetic energy or granular temperature $T$. As expected, the equation for the temperature contains a sink term that accounts for the energy loss in collisions. Less intuitively, the constitutive relation of the heat flux also modifies as \cite{brdukisa98,go03}
\begin{equation*}
  \mn{q}=-\kappa \mn \nabla T-\mu \mn \nabla n.
\end{equation*}
Apart from the term proportional to the temperature gradient ($\kappa$ being the thermal conductivity), another one proportional to the density gradient appears ($\mu$ is known as diffusive conductivity). This new term is purely inelastic in origin, as can be heuristically argued \cite{brru12}. Under most flow conditions, the contribution of the temperature gradient to the heat flux is dominant. However, the term proportional to the density gradient becomes important in regions of the system where the temperature gradient is small enough, situation that usually requires the application of an external force. The presence of the new term in the heat low explains, for example, the non--monotonic temperature dependence on position observed in simulations and experiments \cite{brrumo01,anal16}. In the present work, we identify and fully study the limiting case, namely, a state with homogeneous granular temperature and non--homogeneous density. It turned out that a very specific external position--dependent force, or a uniform external force with a curved two--dimensional silo with a very specific form, is required.

This work is divided into several parts. In the next one, the model of a granular gas in a curved two-dimensional silo is introduced, and equations for three different levels of description are obtained. The main theoretical results of the paper is contained in Sec. \ref{sec3} where we first analyze the steady state of homogeneous temperature and nonzero heat flux within the hydrodynamic Navier--Stokes approach, and then with a mesoscopic one. Contrary to the hydrodynamic results, the general solutions to kinetic equations support infinite one--parametric solutions related to different values of the diffusive conductivity $\mu$. This result is exactly demonstrated for the case of a BGK model \cite{brmodu96}, and approximately using the inelastic Boltzmann equation for hard spheres. The last two sections deal with the comparison between theory and numerical simulations and the conclusions, respectively. The appendixes give extra information about the mathematical calculations and how the numerical simulations were performed, namely event--driven with non--homogeneous external forces.

\section{Model and descriptions \label{sec2}}
\subsection{Model}
\begin{figure}[!h]
  \centering
  \includegraphics[width=.65\textwidth]{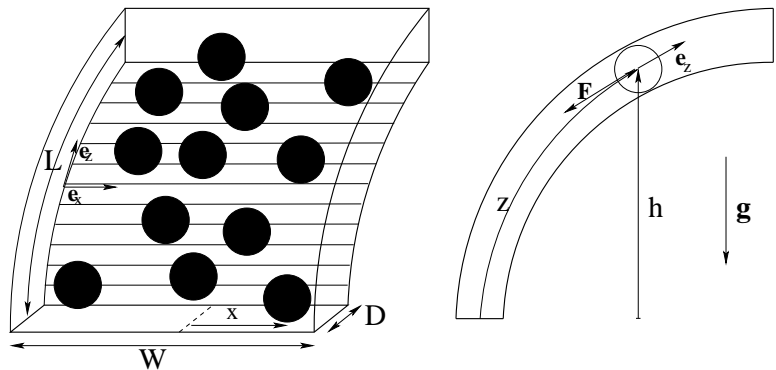}
  \caption{Sketch of the system. The two unitary vectors $\mn{e}_x$ and $\mn{e}_z$ are perpendicular to the lined wall, the first one being constant and horizontal while the second one has a position--dependent direction. The coordinate $z$ measures the distance form the bottom of the system and along the direction of $\mn{e}_z$. The effective force $F$ acting on a grain of mass $m$ is $m\mn{g}\cdot \mn{e}_z$ where $\mn{g}$ is the gravitation acceleration.}
  \label{fig:1}
\end{figure}

The system is modeled as $N$ hard spheres of mass $m$ and diameter $\sigma$ enclosed in a two-dimensional curved silo of length $L$, width $W$, and depth $D$, as shown in Figure \ref{fig:1}. The width of the silo is taken small enough, $D\gtrsim \sigma$, so that the position of a grain $\mn{r}$ is completely determined by the horizontal coordinate $x\in[-W/2,W/2]$ and its distance to the bottom of the silo $z\in[0,L]$. Even more, no friction between grains and the walls of the silo is considered and the gravitation acceleration $\mn{g}$ is taken constant and directed along the vertical direction.

\subsection{Microscopic description}
We refer to the microscopic description of the system as the one that provides the positions and velocities of the grains as a function of time. At this level, the evolution of the system occurs in a sequence of two steps: 
\begin{itemize}
\item [(i)] collision--independent evolution of grains and 
\item [(ii)] instantaneous inelastic collisions among spheres and with the walls.
\end{itemize}
As a function of the generalized coordinates $x$ and $z$, the velocity of a particle can be written as $\mn{v}= \dot{x}\mn{e}_x+\dot{z}\mn{e}_z$, where the dot denotes time derivative, and the unit vectors $\mn{e}_x$ and $\mn{e}_z$ are represented in Figure \ref{fig:1}. The dynamics at stage (i) is governed by two equations:
\begin{eqnarray}
  \label{eq:c1}
  && \ddot{x}=0, \\
  \label{eq:c2}
  && m \ddot{z}= -F(z),
\end{eqnarray}
where $\mn{F}(z)=-F(z) \mn{e}_z$ is the effective external force acting on grains. The amplitude of the force $F(z)$ is given by  
\begin{equation}
  \label{eq:c3}
  F(z)=h'(z)mg,
\end{equation}
where the prime means derivative with respect to $z$. The simplicity of the latter equation is a direct consequence of the definitions of $z$ and $h(z)$. 

The evolution at stage (i) may be interrupted when a grain encounters a wall or another grain. In the latter case, if two grains with velocities $\mn{v}_1$ and $\mn{v}_2$ collide, the post--collision velocities are
\begin{equation}
  \label{eq:c4}
  \mn{v}_1^*=\mn{v}_1-\frac{1}{2}(1+\alpha^{-1})(\mn{v}_{12}\cdot \hnn{\sigma})\hnn{\sigma}, \\
  \mn{v}_2^*=\mn{v}_2+\frac{1}{2}(1+\alpha^{-1})(\mn{v}_{12}\cdot \hnn{\sigma})\hnn{\sigma},
\end{equation}
where $\mn{v}_{12}=\mn{v}_1-\mn{v}_2$ is the relative velocities, $\hnn\sigma$ is a united vector from first to second particle, and $\alpha\in(0,1]$ is the coefficient of normal restitution that tunes the energy loose (the elastic case corresponds to $\alpha=1$). Collisions of particles with the vertical walls are taken elastic, while \emph{thermal} boundary conditions are used for the horizontal ones (see \ref{appen:3}).

We observe that Eqs. (\ref{eq:c1}) and (\ref{eq:c2}) also describe the evolution of grains in a flat silo with external vertical force $-F(z)$. In this case, both $x$ and $z$ are Cartesian coordinates. Moreover, the system of equations can be generalized to describe a $d$--dimensional case where $x$ should be replace by a $(d-1)$--dimensional vector and $z$ coincides with the height $h$. The collision rule (\ref{eq:c4}) does not modify. 


\subsection{Mesoscopic description}

We will assume that for times of the order of the mean free time and distances of the order of the mean free path, the system is well described by the one--distribution function $f(\mn{r},\mn{v},t)$ at time $t$, i.e. the number density of particles around $\mn{r}$ and $\mn{v}$ at time $t$. This probability description of the system is referred to as the mesoscopic one. The balance equation for $f$, since now Boltzmann--like equation, is
\begin{equation}
  \label{eq:c5}
  \frac{\partial}{\partial t} f(\mn{r},\mn{v},t)+\mn{v}\cdot \mn{\nabla}
  f(\mn{r},\mn{v},t)+\frac{\mn{F}}{m} 
  \cdot \mn{\nabla}_{\mn{v}} f(\mn{r},\mn{v},t)=\sigma^{d-1} J[\mn{r},\mn{v},t|f],
\end{equation}
where the l.h.s accounts for the change of $f$ due to stage (i) and the r.h.s accounts for stage (ii). The dimension of the system is $d$. For the selected coordinates, and omitting the functional dependence of $f$ for simplicity, the equation reduces to 
\begin{equation}
  \label{eq:c6}
  \frac{\partial}{\partial t} f+\dot{x} \frac{\partial}{\partial x} f
  +\dot{z} \frac{\partial}{\partial z} f
  -\frac{F(z)}{m} \frac{\partial}{\partial \dot{z}} f=\sigma^{d-1} J[\mn{r},\mn{v},t|f],
\end{equation}
The simple form of the l.h.s. of the equation is a direct consequence of the election of $z$ as the distance from the bottom of the silo. Any other curved coordinate would complicate the expression of the spatial and velocity gradients. As observed in the microscopic description, the equation also corresponds to a system in an euclidean space with a position--dependent external force.

The explicit form of the collision operator $J[\mn{r},\mn{v},t|f]$ depends on the specific model and is not given for now. However, according to the conservation of number of particles and linear momentum in collisions, and since energy is dissipated, the collision operator satisfies
\begin{eqnarray}
  \label{eq:c7}
  && \int d\mn{v} J[\mn{v}_1,\mn{r},t|f]= 0, \\
  \label{eq:c8}
  && \int d\mn{v}\ \mn{v} J[\mn{v}_1,\mn{r},t|f]= 0, \\
  \label{eq:c9}
  && \int d\mn{v} v^2 J[\mn{v}_1,\mn{r},t|f]\le  0.
\end{eqnarray}
The equality on last relation holds only in the elastic case $\alpha=1$. It is worth to remark that the collision operator of the usual Boltzmann equation satisfies Eqs. (\ref{eq:c7})--(\ref{eq:c9}) for almost any function $f$. However, for some other models, additional assumptions, like the normalization of $f$, may be required. 

\subsection{Macroscopic description}
Now we focus on a description that concerns few quantities of the system, namely the number density $n$, the mean velocity $\mn{u}$, and the granular temperature $T$, i.e. the usual hydrodynamic variables, that are defined in terms of the distribution function as
\begin{eqnarray}
  \label{eq:c10}
  n(\mn{r},t)&=&\int d\mn{v} f(\mn{r},\mn{v},t), \\
  \label{eq:c11}
  \mn{u}(\mn{r},t)&=&\frac{1}{n(\mn{r},t)}\int d\mn{v} \mn{v} f(\mn{r},\mn{v},t), \\
  \label{eq:c12}
  T(\mn{r},t)&=&\frac{m}{dn(\mn{r},t)} \int d\mn{v}  (\mn{v}-\mn{u})^2 f(\mn{r},\mn{v},t).
\end{eqnarray}
The corresponding balance equations are obtained by multiplying the Boltzmann--like equation by an appropriate function of the velocity and then integrating it over velocity,
\begin{eqnarray}
  \label{eq:c13}
  &&\frac{\partial}{\partial t} n+\mn{\nabla}\cdot(n\mn{u})=0, \\
  \label{eq:c14}
  &&\frac{\partial}{\partial t} \mn{u}+\mn{u}\cdot \mn{\nabla} \mn{u} +\frac{1}{mn} \mn{\nabla}\cdot P -\frac{\mn{F}}{m}=0, \\
  \label{eq:c15}
    &&\frac{\partial}{\partial t} T+\mn{u}\cdot \mn{\nabla}T+ \frac{2}{dn} P:\mn{\nabla} \mn{u}+\frac{2}{dn} \mn{\nabla}\cdot \mn{q}+ T \zeta=0.
\end{eqnarray}
The pressure tensor $P$, the heat flux $\mn{q}$, and the cooling rate $\zeta$, are also defined in terms of the distribution function,
\begin{eqnarray}
  \label{eq:c16}
  & P_{ij}(\mn{r},t)& = \int d\mn{v} \
  (\mn{v}-\mn{u})_i(\mn{v}-\mn{u})_j f(\mn{r},\mn{v},t), \\
  \label{eq:c17}
  & \mn{q} (\mn{r},t)& =\frac{m}{2} \int d\mn{v} \
  (\mn{v}-\mn{u})^2(\mn{v}-\mn{u}) f(\mn{r},\mn{v},t), \\
  \label{eq:c18}
  & \zeta(\mn{r},t)& =-\frac{m\sigma^{d-1}}{dnT} \int d\mn{v} \ v^2 J[\mn{v}_1,\mn{r},t|f].
\end{eqnarray}

At the first order in the gradients, the Chapman-Enskog method applied to a wide variety of models provides constitutive relations that closes system (\ref{eq:c13})--(\ref{eq:c15}). They read 
\begin{eqnarray}
  \label{eq:c19}
  &&P_{ij} =nT \delta_{ij}-\eta \left(\nabla_i u_j+\nabla_j
    u_i-\frac{2}{d}\delta_{ij} \mn{\nabla}\cdot {\mn{u}} \right), \\
  \label{eq:c20}
  && \mn{q}=-\kappa \mn{\nabla} T- \mu \mn{\nabla}n, \\
  \label{eq:c21}
  && \zeta =\zeta^{(0)},
\end{eqnarray}
with $\eta$, $\kappa$, $\mu$, and $\zeta^{(0)}$ being the transport coefficients, namely the shear viscosity, the thermal conductivity, the diffusive conductivity, and the cooling rate, respectively. They are known functions of the dimensionality $d$, the density $n$, the temperature $T$, and the diameter $\sigma$, mass $m$, and the coefficient of normal restitution $\alpha$ of the grains, the specific functional form being dependent on the corresponding collision operator $J$. The hydrodynamics equations (\ref{eq:c13})--(\ref{eq:c15}) with the constitutive relations (\ref{eq:c19})--(\ref{eq:c21}) are the Navier--Stokes equations. 

\section{State of homogeneous temperature and nonzero heat flux. Theory \label{sec3}}
In this section, a steady state of a granular gas with homogeneous temperature $T=T_0$, zero mean velocity $\mn{u}=0$, and nonzero heat flux is identified and analyzed. First, the hydrodynamic equations (\ref{eq:c13})--(\ref{eq:c15}) with the constitutive equations (\ref{eq:c19})--(\ref{eq:c21}) are considered: one unique solution to the problem is found for only one specific external force $F(z)$. At a mesoscopic scale, the state is identified with a solution to general kinetic Boltzmann-like equations verifying a scaling property. This property leads to the same force $F(z)$ found at the hydrodynamic level, but with infinite different associated hydrodynamic profiles. This unexpected result is exactly demonstrates for a simple BGK model. 

\subsection{Navier--Stokes and $F(z)$}
If the density $n$ depends only on $z$, the temperature $T=T_0$ is uniform, and the mean velocity is zero $\mn{u}=0$, the Navier--Stokes equations can be written as
\begin{eqnarray}
  \label{eq:c22}
  &&\frac{d^2n(s)}{ds^2}=\beta^2_{NS} n(s), \\
  \label{eq:c23}
  && F(s)= -T_0 n(s) \sigma^{d-1} \frac{d}{ds}\ln n(s),
\end{eqnarray}
where the new variable $s$ is defined as
\begin{equation}
  \label{eq:c24}
  s=\int_{0}^z n(z) \sigma^{d-1} dz,
\end{equation}
and the constant $\beta_{NS}$ depends only on $d$ and $\alpha$ through the cooling rate $\zeta$ and the diffusive conductivity $\mu$ as
\begin{eqnarray} 
  \label{eq:c25}
    \beta_{NS}&&=\sqrt{\frac{dT\zeta}{2n^2\sigma^{2(d-1)}\mu}} \\  && \nonumber \simeq \frac{(1+\alpha)\pi^{\frac{d-1}{2}}\sqrt{[4+5d+3(d-4)\alpha][(3+5\alpha)d-8\alpha]}}{\sqrt{4d(d+2)^3}\Gamma\left(\frac{d}{2}\right)}.
\end{eqnarray}
Last approximate equality holds for hard spheres and for $\alpha \sim 1$. Usually it is $\zeta\ge 0$ and $\mu\ge 0$ (the equality signs hold only in the elastic case $\alpha=1$), and hence the relation is well defined. The general solution to Eq. (\ref{eq:c22}) reads
\begin{equation}
  \label{eq:c26}
  n(s)=c_1e^{-\beta_{NS}s}+c_2e^{\beta_{NS}s}
\end{equation}
where $c_1$ and $c_2$ depend on the boundary conditions (which usually involve at the hydrodynamic level the temperature, and the temperature and the density through the heat flux). The force $F(s)$ can be obtained now by inserting $n(s)$ in Eq. (\ref{eq:c23}). However, instead of analyzing the general case, we will focus on ``bulk'' solutions, the ones that are reached by the system far enough from the boundaries. We will suppose since now that $F>0$ and $z\in[0,\infty)$. Far enough from the boundaries, the density profile must be bounded. Moreover, the density must be a decreasing function of $s$, since the force pushes the particles toward the bottom wall. Hence $c_2=0$ and $c_1$ coincides with $n_0$ (the value of the density at $s=0$ or $z=0$),
\begin{equation}
  \label{eq:c27}
  n(s)=n_0e^{-\beta_{NS}s}.
\end{equation}
Now, 
\begin{equation}
  \label{eq:c28}
  F(s)=F_0e^{-\beta_{NS}s},
\end{equation}
with 
\begin{equation}
  \label{eq:c29}
  F_0=\beta_{NS}T_0n_0\sigma^{d-1}.
\end{equation}
This way, the maximum value of the density coincides with the maximum value of the absolute value of the force $F$, since we chose $\mathbf{F}=-F\mathbf{e}_z$. For this ``bulk'' solutions, the variable $s$ is easily expressed as function of $z$, using Eq. (\ref{eq:c27}) in Eq. (\ref{eq:c24}). In doing so, we can rewrite Eqs. (\ref{eq:c27})--(\ref{eq:c29}) as
\begin{eqnarray}
  \label{eq:c30}
  && F(z)=\frac{F_0}{1+z/\lambda}, \\
  \label{eq:c31}
  && n(z)=\frac{1}{\beta_{NS}\lambda\sigma^{d-1}}\frac{1}{1+z/\lambda}, \\
  \label{eq:c32}
  && T_0=F_0 \lambda.
\end{eqnarray}
The constant $\lambda$ is a free parameter that tunes the spatial dependence of the external force. Finally, let $M_{11}$ be the reduced heat flux defied as 
\begin{equation}
  \label{eq:c33} 
  M_{11}=\frac{q_z}{nT_0\sqrt{\frac{2T_0}{m}}},
\end{equation}
If we denote $M_{11}^{NS}$ as the value of $M_{11}$ at the Navier--Stokes order, then using Eqs (\ref{eq:c31}) and (\ref{eq:c32}) in Eq. (\ref{eq:c20}), we obtain 
\begin{equation}
  \label{eq:c34}
  M_{11}^{NS}=\sqrt{\frac{dm\mu\zeta}{4T^2}}\simeq\frac{(1-\alpha^2)\pi^{\frac{d-1}{2}}}{\sqrt{2}\Gamma\left(\frac{d}{2}\right)\beta_{NS}(\alpha)},
\end{equation}
which is zero only in the elastic limit. Last approximate equality holds for hard spheres and for $\alpha \sim 1$.

In conclusion, if our system is subject to the action of the force $F(z)$ in Eq. (\ref{eq:c30}), then the Navier--Stokes equations admit a unique solution, for a given value of the parameters of the system, that corresponds to $\mn{u}=0$, $n(z)$ given by Eq. (\ref{eq:c31}) and a uniform temperature given by Eq. (\ref{eq:c32}). Even more, the heat flux is different from zero and its scaling form is given by $M_{11}^{NS}$ in Eq. (\ref{eq:c34}).

\subsection{Distribution function of the state}
At the kinetic level, and proceeding as in previous works \cite{brcumoru01,brkhru09,brkhdu11,brkhdu12}, we can seek a steady normal solution to the Boltzmann--like equation (\ref{eq:c6}) that corresponds to a bulk solution by imposing the following scaling property to the distribution function
\begin{eqnarray}
  \label{eq:c35}
  f(\mn{r},\mn{v},t)=n(z)v_0^{-d}\phi(\mn{c}),
\end{eqnarray}
with
\begin{eqnarray}
  \label{eq:c36}
  && v_0= \left(\frac{2T_0}{m}\right)^{1/2}, \\
  \label{eq:c37}
  && \mn{c}=\mn{v}/v_0.
\end{eqnarray}
Let $M_{ij}$ be the moments of $\phi$ defined as
\begin{equation}
  \label{eq:c38}
  M_{ij}=\int d\mn{c}\ c^{2i}c_z^j \phi(\mn{c}),
\end{equation}
then, using the definitions of $n$, $\mn{u}$ and $T$ in Eqs. (\ref{eq:c10})--(\ref{eq:c12}) together with the scaling property, we have  the three consistency conditions of $\phi$, 
\begin{eqnarray}
  \label{eq:c39}
  && M_{00}=1,  \qquad M_{01}=0, \qquad M_{10}=\frac{d}{2}.
\end{eqnarray}
Now, substitution of Eq. (\ref{eq:c35}) in Eq. (\ref{eq:c6}) and after some manipulation, the Botlzmann--like equation reduces to
\begin{equation}
  \label{eq:c40}
  -\beta c_z\phi(\mn{c})-\gamma \frac{\partial}{\partial c_z}  \phi(\mn{c})=J'[\mn{c}|\phi], 
\end{equation}
where 
\begin{eqnarray}
  \label{eq:c41}
  J'[\mn{c}|\phi]=\frac{v_0^{d-1}}{n^2} J\left[\mn{r},v_0 \mn{c},t| nv_0^{-d}\phi(\mn{c})\right],
\end{eqnarray}
is the scaled form of the collision operator and 
\begin{eqnarray}
\label{eq:c42}
  && \beta=-\frac{1}{n(z)\sigma^{d-1}} \frac{d \ln n(z)}{dz}, \\
  \label{eq:c43}
  && \gamma=\frac{F(z)}{2T_0 n(z) \sigma^{d-1}}.
\end{eqnarray}
In order to simplify the notation, the prime of $J'$ will be removed and all references to $J$ will refer to the scaled form of $J$. The usual collision operators make $J[\mn{c}|\phi]$ to depend on $z$ only through $\mn{c}$, and hence, for consistency, the l.h.s of Eq. (\ref{eq:c40}) must verify the same property. The most natural possibility, the one to be analyzed here, is that both quantities, $\beta$ and $\gamma$, are constant. 

Upon using the scaling form (\ref{eq:c35}) of the distribution function we are assuming the existence of the hydrodynamic magnitudes. This assumption together with properties (\ref{eq:c7})--(\ref{eq:c9}) of the collision operator can be used to relate constant $\gamma$ with $\beta$ through moment $M_{02}$ of $\phi$. This is accomplished by multiplying Eq. (\ref{eq:c40}) by $c_z$ and integrating over $\mn{c}$, 
\begin{equation}
  \label{eq:c44}
  \gamma=\beta M_{02}[\phi].
\end{equation}
The problem now consists in solving the following equation for $\phi$,
\begin{eqnarray}
  \label{eq:c45}
  -\beta \left\{c_z\phi(\mn{c})+M_{02}[\phi] \frac{\partial}{\partial c_z}  \phi(\mn{c})\right\}=J[\mn{c}|\phi],
\end{eqnarray}
with
\begin{eqnarray}
\label{eq:c46}
\phi \ge 0, \qquad M_{10}=\frac{d}{2}, \qquad (M_{00}=1). 
\end{eqnarray}
The condition $M_{00}=1$ is required only for some collision operators (in order to ensure that the number of particles and linear momentum is conserved in collisions), and the other consistency condition $M_{10}=0$ is automatically fulfilled. One is attempt to remove $\beta$ in favor of some moment $M_{ij}$ by an appropriate integration, but this procedure would be incorrect, since Eq. (\ref{eq:c45}) is a one-dimensional differential equation whose solution has a constant of integration that can be fixed by using condition $M_{10}=d/2$ (and $M_{00}=1$). Another question is whether problem (\ref{eq:c45})--(\ref{eq:c46}) has a solution or not for a given value of $\beta$. This problem will be address bellow. 

Suppose we have managed to solve problem (\ref{eq:c45})--(\ref{eq:c46}) for a given value of $\beta$, at least approximately. Then, integrating Eqs. (\ref{eq:c42}) and (\ref{eq:c43}), we have
\begin{eqnarray}
  \label{eq:c47}
  && F(z)=\frac{F_0}{1+z/\lambda}, \\
  \label{eq:c48}
  && n(z)=\frac{1}{\beta\lambda\sigma^{d-1}}\frac{1}{1+z/\lambda}, \\
  \label{eq:c49}
  && T_0=\frac{1}{2M_{02}[\phi]}F_0 \lambda,
\end{eqnarray}
Finally, the scaled heat flux $M_{11}$ can be obtained by multiplying Eq. (\ref{eq:c45}) by $c^2$ and integrating over $\mn{v}$, with the result
\begin{equation}
  \label{eq:c50}
  M_{11}=-\frac{1}{\beta} \int d\mn{c} c^2 J[\mn{c}|\phi].
\end{equation}

In conclusion, within the kinetic description we obtain the same expression for the force but, in general, different density and temperature profiles in comparison with the hydrodynamic Navier--Stokes description. The results coincide if and only if $\beta=\beta_{NS}$ and $M_{02}[\phi]=1/2$, which is not the case in general since now, on the one hand, $\beta$ is a free parameter and, on the other hand, $M_{02}$ is a function of $\beta$ different from $1/2$. In addition, the heat flux (and its scaling form $M_{11}$) has the same sign as the one of the Navier--Stokes order ($M_{11}^{NS}$), since $\int d\mn{c} c^2 J[\mn{c}|\phi] \le 0$ due to energy dissipation in collisions, but it takes different values.

\subsection*{Exact solution to a BGK model} 
For the BGK model introduced in \cite{brmodu96}, equation (\ref{eq:c45}) with conditions (\ref{eq:c46}) has a unique solution for any $\beta \ge 0$ and any realistic values of the parameters of the system, as we show here and in \ref{appen:1}. Now the scaled collision operator reads 
\begin{equation}
  \label{eq:c51}
  J[\mn{c}|\phi]=-\nu \left[\phi(\mn{c})-\phi_0(\mn{c})\right],
\end{equation}
where $\nu$ is an effective adjustable collision frequency and $\phi_0$ is a reference distribution function:
\begin{equation}
  \label{eq:c52}
  \phi_0(\mn{c})=\pi^{-\frac{d}{2}}\alpha^{-d}\exp\left(-\frac{c^2}{\alpha^2}\right).
\end{equation}
As is proven in \ref{appen:1}, the solution is
\begin{equation}
  \label{eq:c53}
  \phi(\mn{c})=\pi^{-\frac{d-1}{2}}\alpha^{1-d}\exp\left(-\frac{c^2-c_z^2}{\alpha^2}\right) \phi_z(c_z),
\end{equation}
with
\begin{equation}
  \label{eq:c54}
  \phi_z(c_z)=\left[A+B(c_z)\right] \exp\left(-\frac{c_z^2}{2M_{02}}+\frac{\nu}{\beta M_{02}} c_z\right).
\end{equation}
In last expression, $A$ is a constant of integration that comes form the general solution to the homogeneous equation associated to Eq. (\ref{eq:c45}) and $B(c_z)$ comes form a particular solution and is given by 
\begin{equation}
  \label{eq:c55}
  B(c_z)=\pi^{-\frac{1}{2}} \alpha^{-1} \frac{\nu}{\beta M_{02}} \int_{c_z}^\infty dx \ \exp\left[-\left(\frac{1}{\alpha^2}-\frac{1}{2M_{02}}\right)x^2-\frac{\nu}{\beta M_{02}} x\right].
\end{equation}
Constant $A$ is determined by imposing the normalization condition in \ref{appen:1}. In addition, we prove that it is positive for any values of the parameters, and hence that $\phi\ge 0$. In addition, we have
\begin{equation}
  \label{eq:c56}
  M_{02}=\frac{1}{2} +\frac{d-1}{2}(1-\alpha^2).
\end{equation}
With this last relation, we obtain the hydrodynamic profiles and, using again the scaling property (\ref{eq:c35}), the distribution function. Finally, the scaled heat flux is
\begin{equation}
  \label{eq:c57}
  M_{11}=\frac{d(1-\alpha^2)\nu}{2\beta}.
\end{equation}
Constant $\nu$ can be now fixed, for instance, by imposing $M_{11}=M_{11}^{NS}$ for $\beta=\beta_{NS}$.

\subsection*{Sonine approximation}
\begin{figure}[!h]
\begin{center}
\includegraphics[width=.475\textwidth,angle=0]{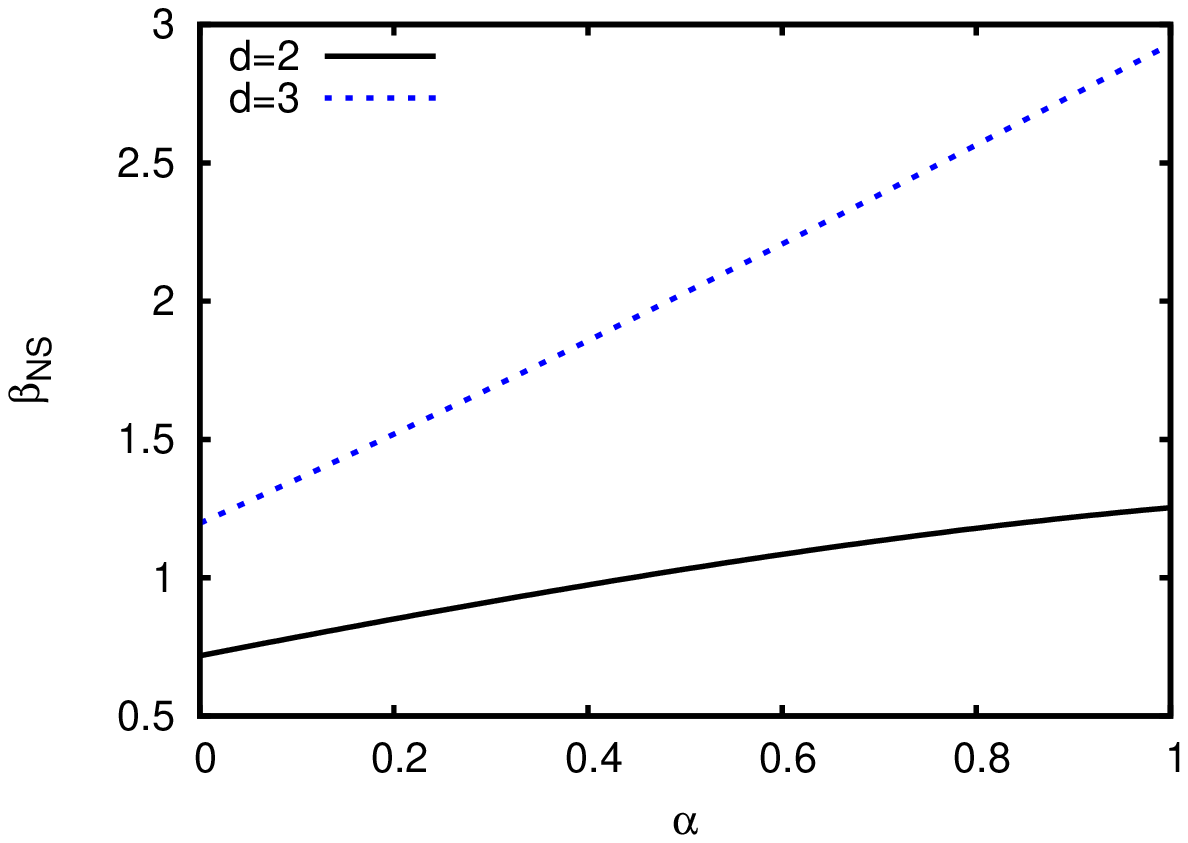}
\includegraphics[width=.475\textwidth,angle=0]{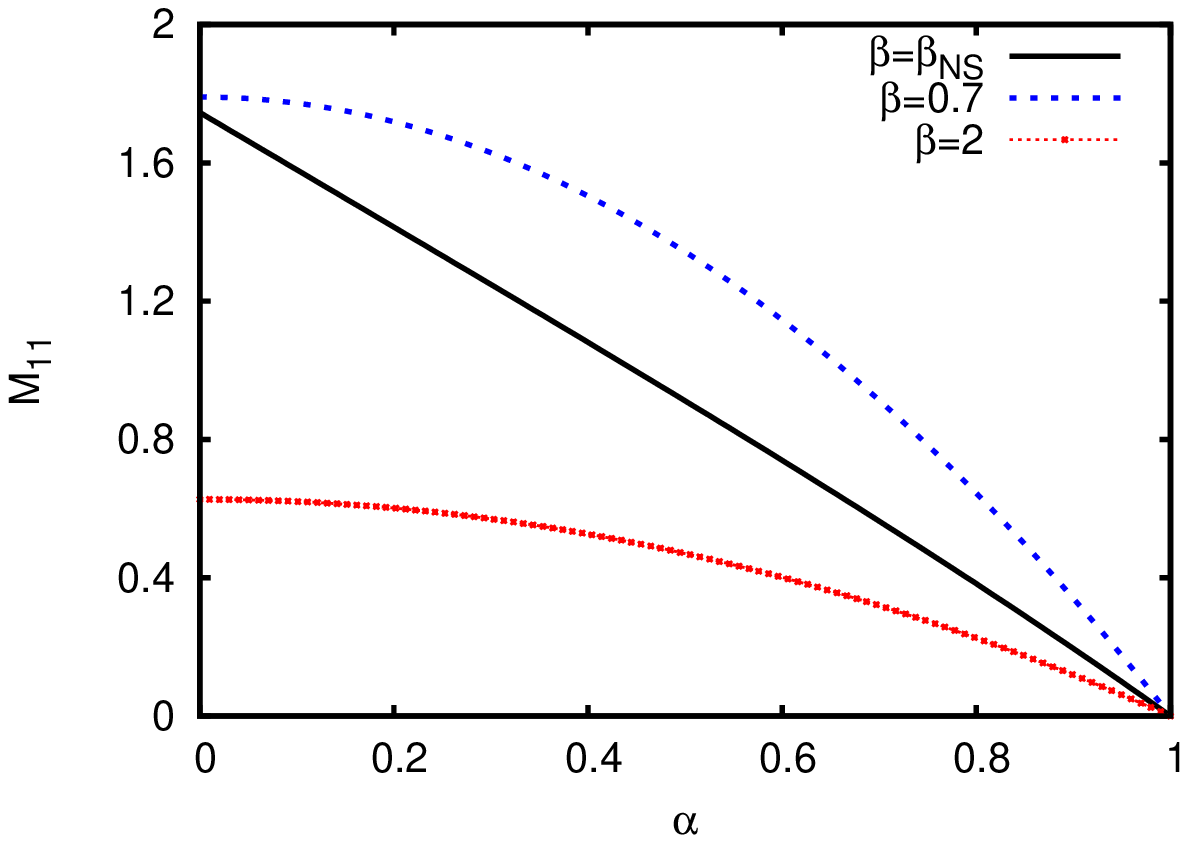}
\caption{Left: coefficient $\beta_{NS}$ given in Eq. (\ref{eq:c25}) for $d=2$ and $d=3$ as a function of $\alpha$. Right: scaled heat flux $M_{11}$, given in Eq. (\ref{eq:c60}), as a function of $\alpha$ and three different values of $\beta$.\label{fig:2}}
\end{center}
\end{figure}

For the Boltzmann collision operator, and according to the considerations made in \ref{appen:2}, the distribution function is approximated as
\begin{eqnarray}
  \label{eq:c58}
  \phi(\mn{c})\simeq && \pi^{-d/2} e^{-c^2} \Bigg[1-a_{01}(c^2-dc_x^2)+ \\
    && \nonumber   +\left(\frac{d-1}{2}b_{01} +\frac{3}{2}
      b_{10}\right) c_x -b_{01}c^2c_x-(b_{10}-b_{01})c_x^3\Bigg].
 \end{eqnarray}
where $a_{01}$, $b_{10}$, and $b_{01}$ are approximately evaluated in \ref{appen:2}. For $(1-\alpha)/\beta \ll 1$, see Figure \ref{fig:2}, we obtain
\begin{equation}
  \label{eq:c59}
  M_{02}=\frac{1}{2}+\frac{d-1}{4d}(1-\alpha^2)
\end{equation}
and 
\begin{equation}
\label{eq:c60}
  M_{11}=\frac{(1-\alpha^2)\pi^{\frac{d-1}{2}}}{\sqrt{2}\Gamma\left(\frac{d}{2}\right)\beta}
\end{equation}
which coincides with Eq. (\ref{eq:c34}) of the Chapman-Enskog  method, at the same Sonine approximation, only for $\beta=\beta_{NS}$. 

\section{State of homogeneous temperature and nonzero heat flux. Simulations \label{sec4}}

\subsection{Molecular dynamics simulations}

A two-dimensional system has been simulated by means of a modified event--driven molecular dynamics algorithm, as explained in \ref{appen:3}. The system has $N$ particles with diameter $\sigma$ and mass $m$, chosen to be units of length and mass respectively. A position-dependent external force $F$, given by Eq. (\ref{eq:c47}), acts on each particle. The absolute value of the force $F_0$ at $z=0$ is used as unit of force, and together with $\sigma$ and $m$ fixes the unit of time. 

We run three sequences of simulations, for $\beta\simeq \beta_{NS}/2, \beta_{NS}, 2 \beta_{NS}$ and different values of the dissipation (coefficient $\alpha$). In order to reach the state with homogeneous temperature and inverse linear density profile, with different values of $\beta$, we used thermal boundary conditions, as explained in \ref{appen:3}, where the partial temperatures of the rebounded particles where fixed. For all simulations, we took the values $W=200\sigma$, $L=400\sigma$, $F_0=1$, and $\lambda=100\sigma$.

In order to minimize the boundary effects and have a wide ``bulk'', the $\alpha$'s were always close to $1$ because of two reasons:
\begin{itemize}
\item[(i)] Variable $s_m$, the maximum value of $s$ defined in Eq. (\ref{eq:c30}), and proportional to the total mean free paths of the system, must be big enough. It is given by 
\begin{equation}
  \label{eq:68}
  s_m=\frac{1}{\beta}\ln \left(1+\frac{L}{\lambda}\right).
\end{equation}
\item[(ii)] The scaled distribution functions of the rebounded particles must be similar to the scaled distribution function far from the boundaries, according to the scaling property in Eq. (\ref{eq:c35}). With the thermal boundary conditions used in the simulations, see \ref{appen:3}, this is expected to occur if the coefficients $b_{01}$ and $b_{10}$ are small enough, i.e. for $(1-\alpha^2)/\beta$ small enough.
\end{itemize}
That is, for given values of $L$ and $\lambda$, $\beta$ must be small, in order to be $s_m$ big. On the contrary, $(1-\alpha^2)/\beta$ must be small, and hence, $\alpha$ close to $1$.

\subsection{Results}
Here, we compare the more relevant theoretical predictions, namely the density and temperature profiles and the scaled heat flux, with the numerical simulation results. 

\begin{figure}[!h]
\begin{center}
\includegraphics[width=.475\textwidth,angle=0]{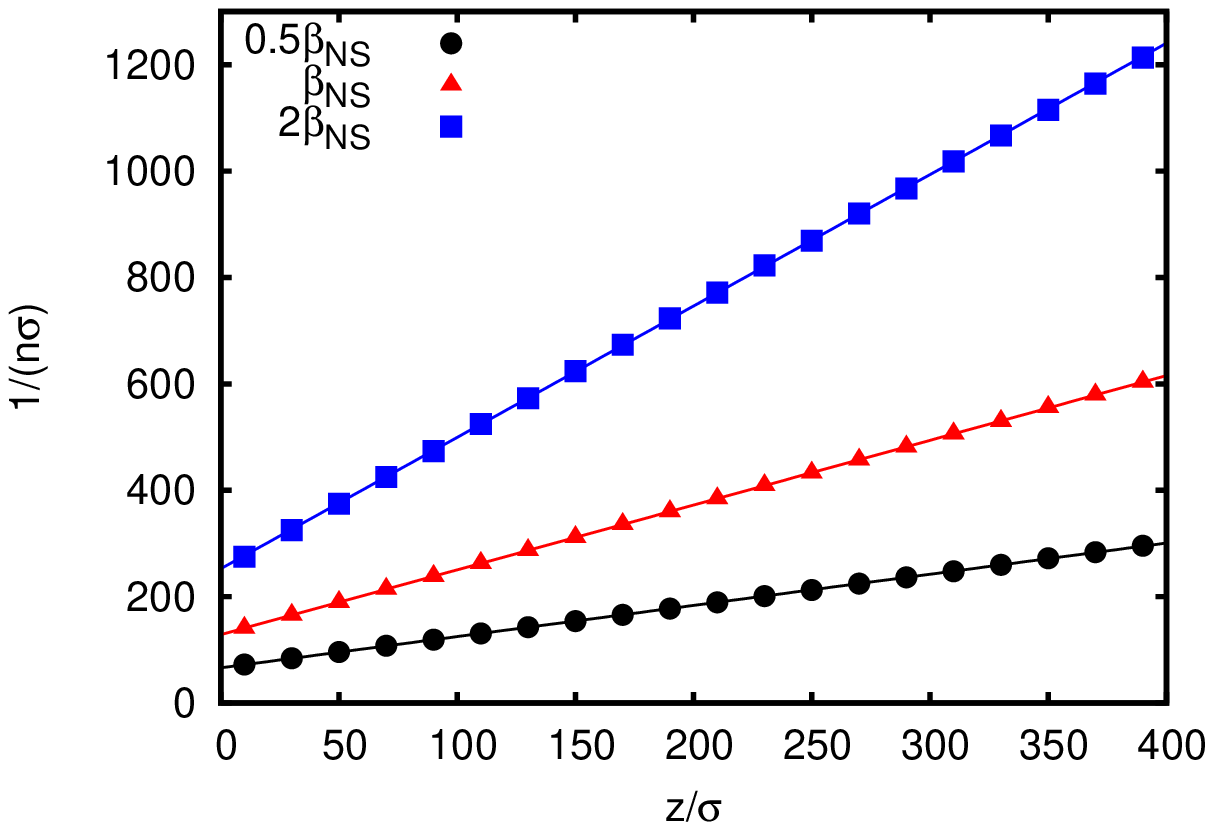}
\includegraphics[width=.475\textwidth,angle=0]{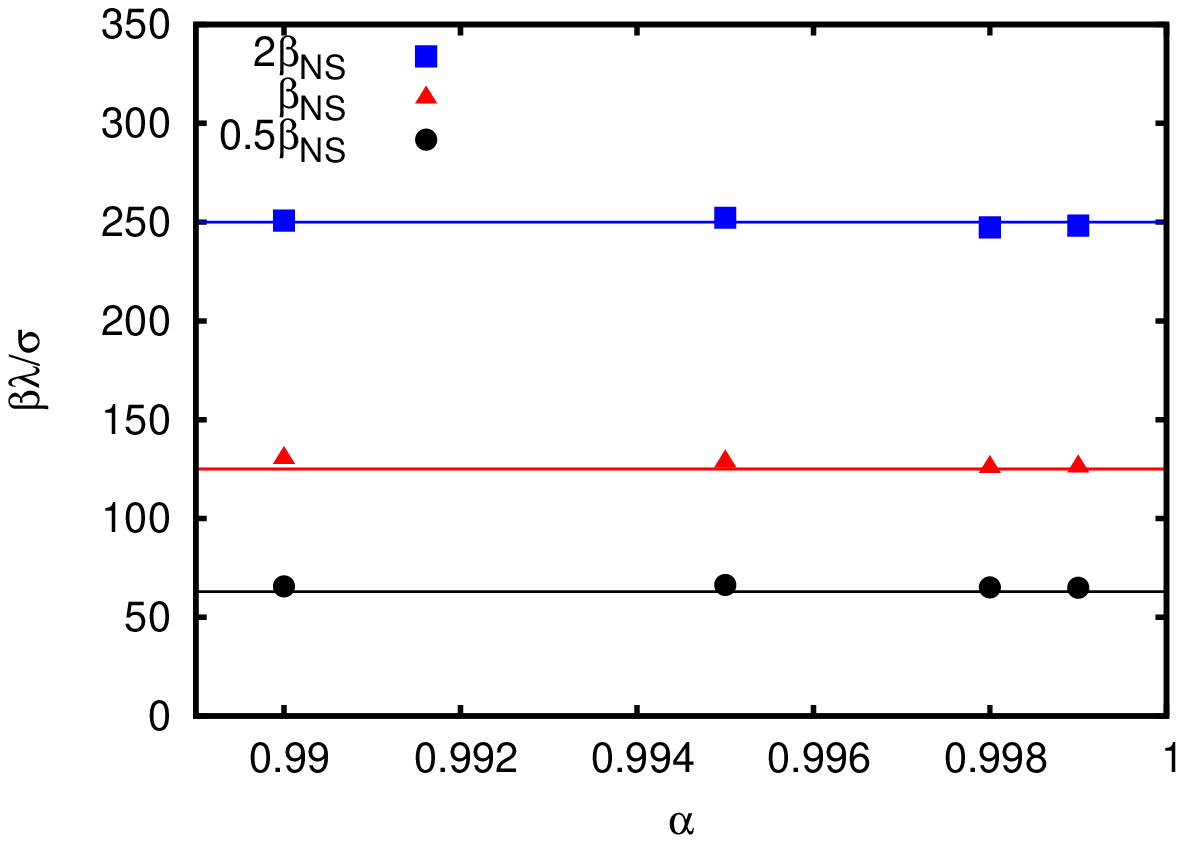}
\caption{Left: the symbols correspond to the inverse of the density as a function of $z$ for three different values of $\beta$ and for $\alpha=0.995$. Lines correspond to best linear fits. Right: symbols represent values of $\beta \lambda$ extracted form the best fists of the left plot as a function of $\alpha$ and for different $\beta$'s. Lines are theoretical values. \label{fig:3}}
\end{center}
\end{figure}

Equations (\ref{eq:c31}) and (\ref{eq:c48}) predict a linear behaviour of the inverse of the number density $1/n$ as a function of the distance $z$. This result is corroborated by simulations, as the left plot of Figure \ref{fig:3} shows. Not only a linear profile of $1/n$ is observed for $\beta=\beta_{NS}$, according to the hydrodynamic Navier--Stokes description, but also for other $\beta$'s, as Kinetic Theory predicts. The right plot of Figure \ref{fig:3} shows the values of $\beta \lambda$ obtained through the linear fittings of $1/n$, with symbols, and its respective theoretical values using Eq. (\ref{eq:c48}), with lines. Since $\beta \lambda$ gives the theoretical value of the number density at the bottom of the system, see Eq. (\ref{eq:c48}), the deviation of theory and simulations shown in the right plot of Fig. \ref{fig:3} is a measure of the width of the bulk.

\begin{figure}[!h]
\begin{center}
\includegraphics[width=.475\textwidth,angle=0]{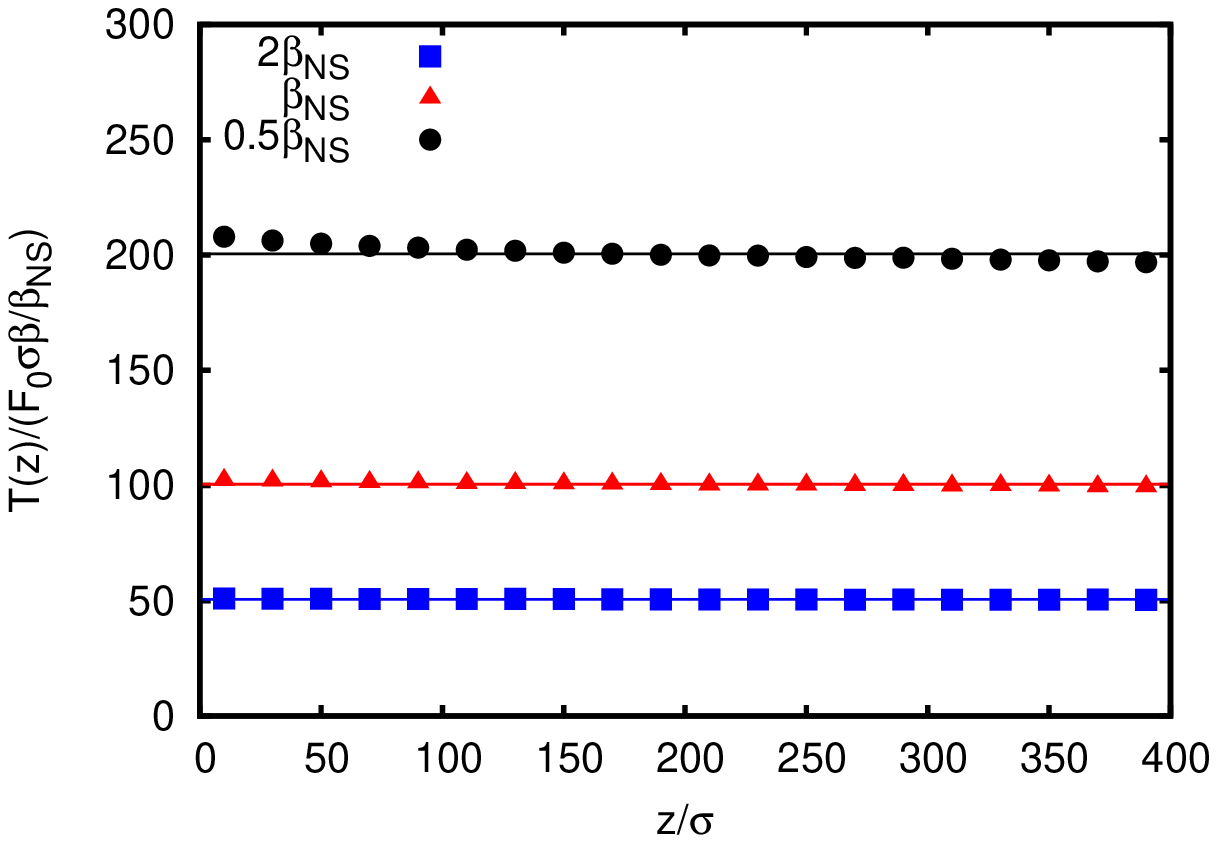}
\includegraphics[width=.475\textwidth,angle=0]{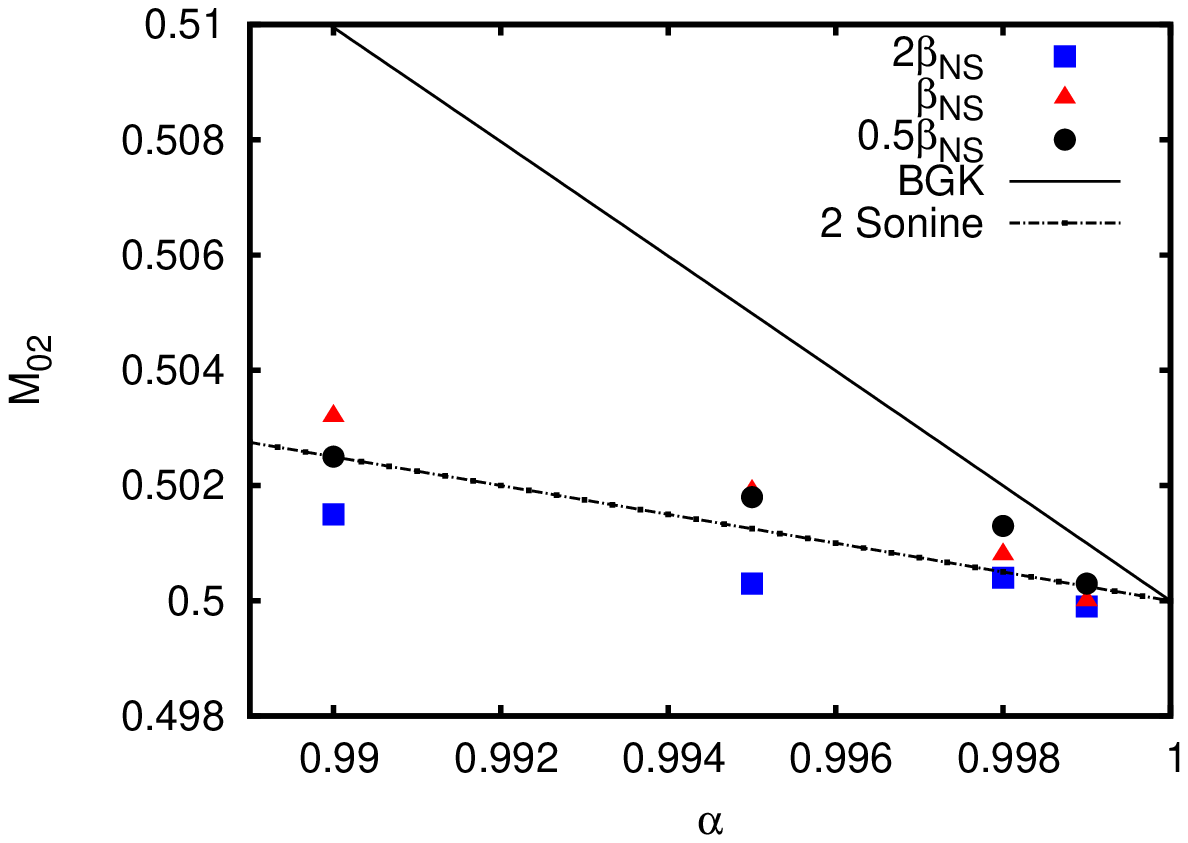}
\caption{Left: non--dimensional temperature $T(z)/(F_0 \sigma)$ divided by $\beta/\beta_{NS}$ as a function of $z$ and for $\alpha=0.995$. Lines correspond to best fits. Right: $M_{02}$ as a function of $\alpha$ and different parameters of the system. Lines are the theoretical predictions of the BGK model, Eq. (\ref{eq:c56}), and the Sonine approximation.\label{fig:4}}
\end{center}
\end{figure}

Another important prediction of the Navier--Stokes and kinetic descriptions is a flat temperature profile given by Eqs. (\ref{eq:c32}) and (\ref{eq:c49}), respectively. Left plot of Fig. \ref{fig:4} shows the temperature measured at different values of $z$ (symbols) and best fits to a constant function for different values of $\beta$ and for $\alpha=0.995$. The values of the temperature $T$ has been divided by $\beta/\beta_{NS}$ in order to avoid an interference among data. The observed values are very close to the theoretical one, i.e $T_0=100 F_0 \sigma$ for all of them. However, the Kinetic Theory analysis predicts slight deviations with respect to the latter value because of the dependence of $M_{02}$ on $\alpha$. But $M_{02}$ is very close to $1/2$ as shows the right plot of Fig. \ref{fig:4}, where we represent the measured values of $M_{02}$ (symbols) averaged all over the system as a function of $\alpha$, as well as the theoretical predictions of the BGK model and the Sonine approximation to the Botlzmann operator (lines).

\begin{figure}[!h]
\begin{center}
\includegraphics[width=.475\textwidth,angle=0]{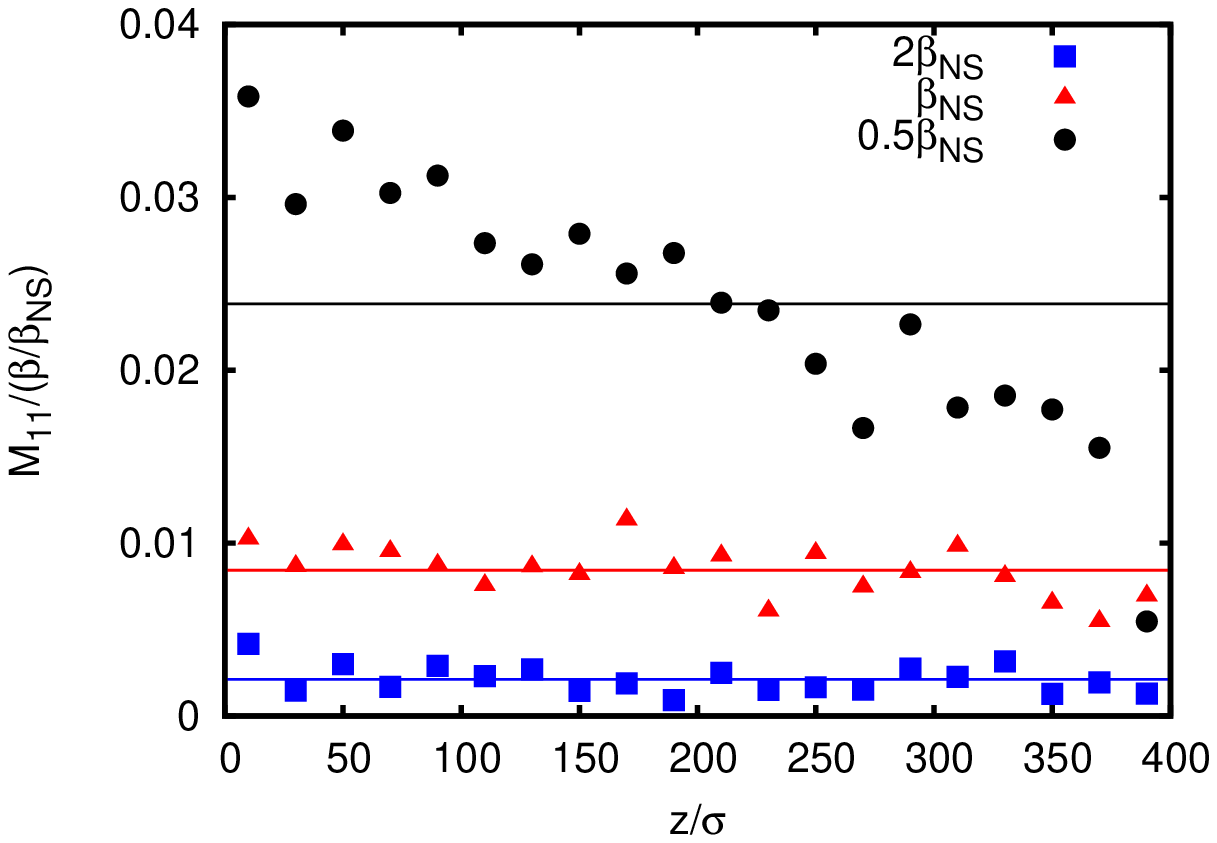}
\includegraphics[width=.475\textwidth,angle=0]{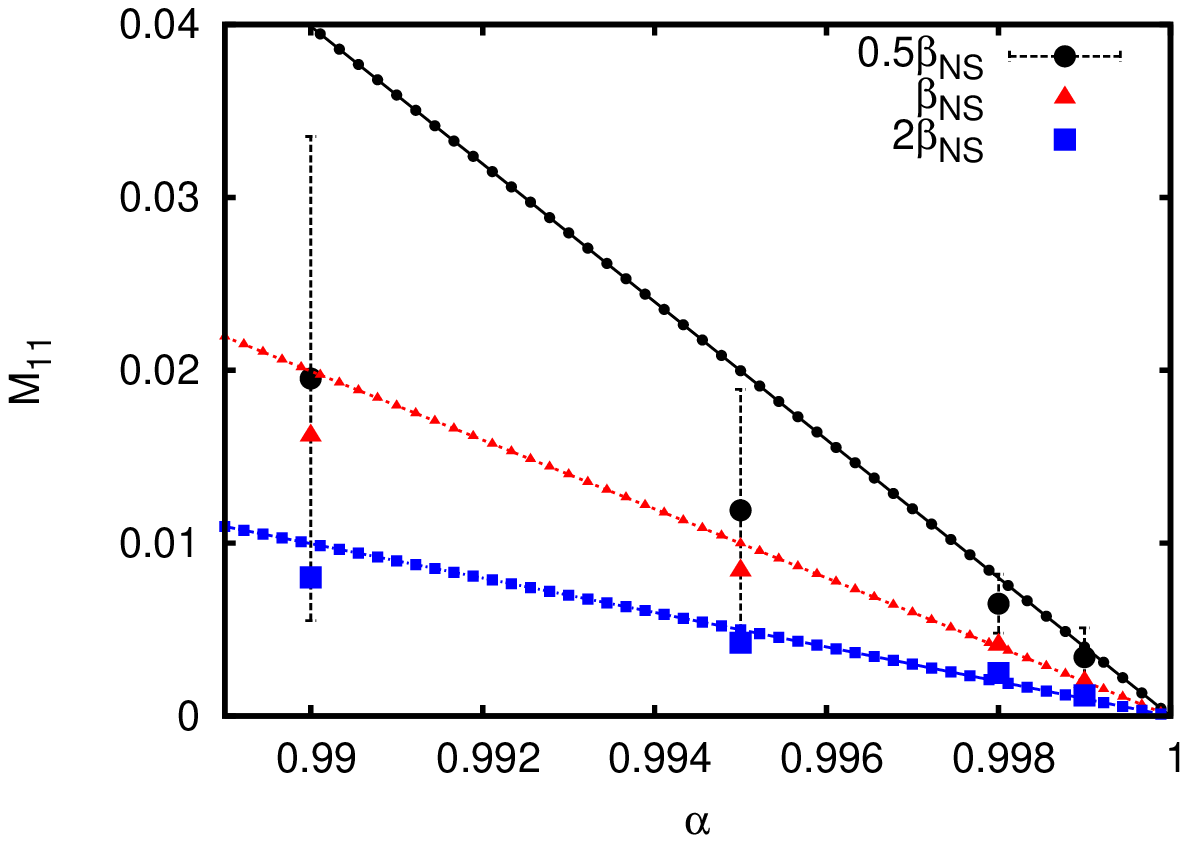}
\caption{Left: scaled heat flux $M_{11}$ divided by  $\beta/\beta_{NS}$ as a function of $z$ for three values of $\beta$ and for $\alpha=0.995$. Lines correspond to best fits. Right: scaled heat flux $M_{11}$ averaged over the system (symbols) and predictions of Eq. (\ref{eq:c60}) (lines) as a function of $\alpha$ and for three different values of $\beta$.\label{fig:5}}
\end{center}
\end{figure}

Finally, the scaled heat flux of the state is nonzero and constant along the system. This property is clearly appreciated in the left plot of Fig. \ref{fig:5} for three values of $\beta$ and $\alpha=0.995$. The theoretical predictions and simulations results are shown in the right plot of Fig. \ref{fig:5}. Now, deviations are apparent since bigger moments of the distribution function are more sensitive to the effect of the boundaries.

\section{Discussion and conclusions }

In this work a steady state of a granular gas with homogeneous granular temperature, zero mean flow, and nonzero heat flux was identified and studied. The study has been carried out using three complementary approaches: hydrodynamics at Navier--Stokes order, Kinetic Theory, and molecular dynamics simulations. The analysis of the Navier--Stokes equations for the granular gas reveals that for the system to reach the desired state, an external position--dependent force directed along one direction of the system is required. Moreover, the state corresponds to a bulk one, in the sense that it may be reached by the system far enough from the boundaries (provided hydrodynamic boundary conditions are used). Since it is experimentally difficult to have a non--homogeneous force acting on the grains without introducing forces among grains, we propose a way to \emph{create} the external force by applying an homogeneous external force (gravitation) and making the grains live inside a curved two--dimensional silo. A general position dependent forces can be generated by tuning the form of the silo. It is worth to mention that the possible frictional force among grains and walls are removed, otherwise other approaches like the ones in \cite{pulomapevu98,gachve13,gachve13a,khga13} may be required. Moreover, the energy injection needed for the steady state to be reached by the system is done through walls on the bottom and top of the system \cite{anal16}, and hence the confined system is not globally fluidizid like in recent studies \cite{ripogarisoco11,camuso12,brriso13,soribr14,brbugama16}.

As the usual Navier--Stokes description of the granular gas is based on a particular way of applying the Chapman-Enskog method to solve the inelastic Boltzmann equation, and hence its applicability is reduced, we also addressed the direct study of the Botlzmann--like equation. The results of the Navier--Stokes equations is completely different from the corresponding mesoscopic description: for a given value of the parameters of the system and for the same form of the external force or shape of the silo, the hydrodynamics supports only one possible values of the macroscopic profiles, given by Eqs. (\ref{eq:c31})--(\ref{eq:c33}), while the kinetic equations provide infinite number of them, now given by Eqs. (\ref{eq:c48})--(\ref{eq:c50}). The way the system chooses one of the possible solutions is not only determined by values of the hydrodynamic magnitudes at the boundaries, but also through the specific form of the distribution function. In other words, with the same external force or shape of the silo, we can have infinite situations where the temperature is uniform, the inverse of the number density is linear with $z$, an the scaled distribution function is the same all over the system. 

In order to give more support to the theoretical results of the general formulation, we considered not only the usual way of addressing the problem, consisting in solving the Boltzmann equation approximately, but also a BGK model and event--driven simulations. On the one hand, for the BGK model, we obtained exact and explicit results for all the macroscopic profiles and the heat flux, Eqs (\ref{eq:c56}) and (\ref{eq:c57}), and also for the distribution function, Eqs. (\ref{eq:c53}) and (\ref{eq:c54}). On the other hand, the numerical simulations of the present work confirm the theoretical predictions at least for weak dissipation, namely $\alpha \gtrsim 0.99$. Apart from the important limitation of the simulations, discussed in Sec. \ref{sec4}, there is another one related to the nature of our state. Actually, the state we have studied has the particularity of a bulk solution, and for that, the distribution function obeys a scaling form for any position. If we generate at the boundaries a distribution function that is close, but not equal, to the desired one, the system takes more space to transform it into the correct one, since the nature of the state is to keep the scaled distribution function as it is, the boundary layers being bigger and more difficult to remove in this case than in other states.

Finally, it is worth to emphasize that the results of the present work seem to reflect an important limitation of the usual inelastic Navier--Stokes equations, not only when the system has shear \cite{ga06,lu06,sagadu04,brrumo97}, but also when nonuniform external forces are applied. 

\ack The author acknowledges partial financial support from Ministerio de Econom\'ia y Competitividad and Fondo Europeo de Desarrollo Regional under project FIS2015-63628-C2-2-R (MINECO/FEDER).

\appendix

\section{BGK model \label{appen:1}}
Equation (\ref{eq:c45}) with conditions (\ref{eq:c46}) can be exactly solved for a BGK model. In this case the scaled collision operator is given by Eq. (\ref{eq:c51}) and (\ref{eq:c52}). The general solution to the resulting Boltzmann equation can be factorized as in Eq. (\ref{eq:c53}). 

The factorization property of the solution has formidable consequences. If used with the normalization condition $M_{00}=1$, imposes the normalization condition to the marginal distribution $\phi_z$,
\begin{eqnarray}
  \label{eq:apa1}
  &&\int_{-\infty}^\infty dc_z \ \phi_z(c_z)=1.
\end{eqnarray}
If used with condition $M_{10}=\frac{d}{2}$, it provides Eq. (\ref{eq:c56}), i.e. $M_{02}$ becomes a known function $d$ and $\alpha$. Finally, if used with the Botlzmann equation, it gives the following problem
\begin{eqnarray}
  \label{eq:apa2}
  -\beta \left\{c_z\phi_z(c_z)+M_{02} \frac{d}{d c_z}
  \phi_z(c_z)\right\}=-\nu(\phi_z-\phi_{0z}),
\end{eqnarray}
with
\begin{eqnarray}
\label{eq:apa3}
\phi_z \ge 0, \qquad   \int_{-\infty}^\infty dc_z \ \phi_z(c_z)=1,
\end{eqnarray}
where
\begin{equation}
  \label{eq:apa4}
  \phi_{0z}(c_z)=\pi^{-\frac{1}{2}}\alpha^{-1}\exp\left(-\frac{c_z^2}{\alpha^2}\right).
\end{equation}

The general solution to Eq. (\ref{eq:apa2}) can be written as in Eq. (\ref{eq:c54}) that depends on a constant $A$ and a function $B(c_z)$, Eq. (\ref{eq:c55}). Function $B(c_z)$ is well defined since for any values of the parameters, using Eq. (\ref{eq:c56}),  it is
\begin{equation}
  \label{eq:apa5}
  \left(\frac{1}{\alpha^2}-\frac{1}{2M_{02}}\right)=\frac{d(1-\alpha^2)}{2M_{02}\alpha^2} \ge 0.
\end{equation}
The constant $A$ should be chosen in order to ensure conditions (\ref{eq:apa3}). From Eq. (\ref{eq:c55}), it is
\begin{eqnarray}
  \label{eq:apa6}
   0\le B(c_z)\le && \pi^{-\frac{1}{2}} \alpha^{-1} \frac{\nu}{\beta M_{02}} \int_{c_z}^\infty dx \ \exp\left(-\frac{\nu}{\beta M_{02}} x\right) \\ && =\pi^{-\frac{1}{2}} \alpha^{-1} \exp\left(-\frac{\nu}{\beta M_{02}} c_z\right)
  \stackrel{c_z\to \infty}{\longrightarrow} 0  \nonumber,
\end{eqnarray}
and hence $\phi_z$ is positive for any $c_z$ if and only if 
\begin{equation}
  \label{eq:apa7}
 A\ge 0. 
\end{equation}

From the normalization condition of $\phi_z$, it is 
\begin{equation}
  \label{eq:apa8}
  A=\frac{1-\mathcal I}{\int_{-\infty}^\infty dc_z\ \exp\left(-\frac{c_z^2}{2M_{02}}+\frac{\nu}{\beta M_{02}}c_z \right)}.
\end{equation}
where 
\begin{equation}
  \label{eq:apa9}
  \mathcal{I}=\int_{-\infty}^\infty dc_z \ B(c_z) \exp\left(-\frac{c_z^2}{2M_{02}}+\frac{\nu}{\beta M_{02}}c_z \right).
\end{equation}
It can be proven that $A\ge 0$, or equivalently, that $\mathcal I \le 1$. Namely, on the one hand, $\mathcal I$ can be written as 
\begin{equation}
  \label{eq:apa10}
\mathcal I=\frac{\beta M_{02}}{\nu}\int_{-\infty}^\infty dc_z \ B(c_z) \exp\left(-\frac{c_z^2}{2M_{02}}\right)
\frac{d}{dc_z}\exp\left(\frac{\nu}{\beta M_{02}}c_z \right),
\end{equation}
and, after an integration by parts, $\mathcal I$ is 
\begin{eqnarray}
  \label{eq:apa11}
  \mathcal I =  1+ \frac{1}{\alpha M_{02} \sqrt{\pi}} \int_{-\infty}^\infty  && dc_z \exp\left(-\frac{c_z^2}{2M_{02}}+\frac{\nu}{\beta M_{02}}c_z \right) \\ && \times \int_{c_z}^\infty dx \ c_z \exp\left[-\frac{d(1-\alpha^2)}{2M_{02}\alpha^2}x^2-\frac{\nu}{\beta M_{02}} x\right] . \nonumber 
\end{eqnarray}
On the other hand, $B(c_z)$ can also be written as 
\begin{equation}
  \label{eq:apa12}
  B(c_z)=-\frac{1}{\alpha\sqrt{\pi}}\int_{c_z}^\infty dx \ \exp\left[-\frac{d(1-\alpha^2)}{2M_{02}\alpha^2}x^2 \right] \frac{d}{dx}\exp \left[-\frac{\nu}{\beta M_{02}} x\right],
\end{equation}
and, after an integration by part and an insertion in Eq. (\ref{eq:apa10}), $\mathcal{I}$ becomes 
\begin{eqnarray}
  \label{eq:apa13}
  \mathcal I =  1- \frac{d(1-\alpha^2)}{\alpha M_{02} \sqrt{\pi}} \int_{-\infty}^\infty  && dc_z \exp\left(-\frac{c_z^2}{2M_{02}}+\frac{\nu}{\beta M_{02}}c_z \right) \\ && \times \int_{c_z}^\infty dx \ x \exp\left[-\frac{d(1-\alpha^2)}{2M_{02}\alpha^2}x^2-\frac{\nu}{\beta M_{02}} x\right] . \nonumber 
\end{eqnarray}
Operating on Eqs. (\ref{eq:apa11}) and (\ref{eq:apa13}), it is
\begin{eqnarray}
  \label{eq:apa14}
  2\sqrt{\pi}M_{02} && \frac{1+d(1-\alpha^2)}{d(1-\alpha^2)} (\mathcal I -1)  =  
  - \int_{-\infty}^\infty dc_z    \exp\left(-\frac{c_z^2}{2M_{02}}+\frac{\nu c_z}{\beta M_{02}} \right) \\
  &&  \times \int_{c_z}^\infty dx \ (x-c_z) \exp\left[-\frac{d(1-\alpha^2)}{2M_{02}\alpha^2}x^2-\frac{\nu}{\beta M_{02}} x\right]. \nonumber
\end{eqnarray}
The rhs of last equation is negative and, since the coefficient multiplying $\mathcal I -1$ on the lhs is positive, it is 
\begin{equation}
  \label{eq:apa15}
  \mathcal I \le 1.
\end{equation}

\section{Sonine expansion \label{appen:2}}
In this appendix, we give mathematical support to expression (\ref{eq:c58}) as an approximate solution to problem (\ref{eq:c45})--(\ref{eq:c46}). 

\subsection*{Space of functions with finite moments}
Let $L^2(\Re^d)$ be the Hilbert space of square integrable real functions of $\mn{c}\in\Re^d$ with the usual scalar product denoted by $\braket{}{}_0$. The vector space
\begin{equation}
  \label{eq:apb1}
  \mathcal{M}=\{f(\mn{c})=e^{-\frac{c^2}{2}}g(\mn{c})\  | \ g(\mn{c})\in L^2(\Re^d) \ even \ on\ \mn{c}_\perp \}
\end{equation}
contains only even functions of $\mn{c}_\perp=\mn{c}-c_z\mn{u}_z$ and is also a Hilbert space with the scalar product defined as
\begin{eqnarray}
  \label{eq:apb2}
  \braket{f}{g}=\int d\mn{c} e^{c^2} f(\mn{c}) g(\mn{c}).
\end{eqnarray}
The new Hilbert space contains functions with finite moments $M_{ij}$, as defined in Eq. (\ref{eq:c38}). If $f\in\mathcal{M}$ then $f=e^{-\frac{c^2}{2}}g$ with $g\in L^2(\Re^d)$ and 
\begin{equation}
  \label{eq:apb3}
  M_{ij}=\int d\mn{c} c^{2i}c_z^j f(\mn{c})=\braket{e^{-\frac{c^2}{2}}c^{2i}c_z^j}{g}_0<\infty
\end{equation}
since $e^{-\frac{c^2}{2}}c^{2i}c_z^j\in L^2(\Re^d)$. It is possible to construct a bases of $\mathcal{M}$ using the Sonine Polynomials. The latter can be defined as 
\begin{equation}
\label{eq:apb4}
S_{n}^{(i)}(v^2) = \sum_{k=0}^{i} \frac{(-1)^{k} \Gamma (n+i+1)}{\Gamma (n+k+1)(i-k)! k!}\, v^{2k}, \quad \mn{v}\in \Re^D, 
\end{equation}
and verify 
\begin{equation}
\label{eq:apb5}
\int d\mn{v}\, v^{2n-(D-2)} e^{-v^2} S_{n}^{(i)} (v^2) S_{n}^{(j)}(v^2) = \frac{\pi^{D/2}\Gamma (n+i+1)}{i!\Gamma\left(\frac{D}{2}\right)}\, \delta_{ij}, \quad \mn{v}\in \Re^D,
\end{equation}
where $D$ is any positive natural number and $\delta_{ij}$ the Kronecker delta. Now it is straightforward to verify that $\mathcal{B}=\{E_{ij}(\mn{c}),O_{ij}(\mn{c})\}_{i,j}$, where $i$ and $j$ natural numbers range from $0$ to $\infty$, and
\begin{eqnarray}
  \label{eq:apb6}
  && E_{ij}(\mn{c})=\sqrt{\pi^{-\frac{d}{2}} \frac{i!j! \Gamma\left(\frac{1}{2}\right) \Gamma\left(\frac{d-1}{2}\right)}{\Gamma\left(i+\frac{1}{2}\right) \Gamma\left(j+\frac{d-1}{2}\right)}} e^{-c^{2}} S_{-\frac{1}{2}}^{(i)} (c_{z}^{2})S_{\frac{d-3}{2}}^{(j)} (c_{\perp}^{2}), \\
  \label{eq:apb7}
  && O_{ij}(\mn{c})=\sqrt{\pi^{-\frac{d}{2}} \frac{i!j! \Gamma\left(\frac{1}{2}\right) \Gamma\left(\frac{d-1}{2}\right)}{\Gamma\left(i+\frac{3}{2}\right) \Gamma\left(j+\frac{d-1}{2}\right)}} e^{-c^{2}} S_{\frac{1}{2}}^{(i)} (c_{z}^{2})S_{\frac{d-3}{2}}^{(j)}(c_{\perp}^{2})c_z,
\end{eqnarray}
is an orthonormal (and complete) set on $\mathcal{M}$ with the scalar product (\ref{eq:apb2}). Therefore, any $f\in \mathcal{M}$ can be expanded as
\begin{equation}
  \label{eq:apb8}
  f(\mn{c})=\sum_{i,j} \left(a_{ij} E_{ij}(\mn{c})+b_{ij}O_{ij}(\mn{c})\right),
\end{equation}
with coefficients $a_{ij}$ and $b_{ij}$ given by 
\begin{eqnarray}
  \label{eq:apb9}
  && a_{ij}= \braket{f}{E_{ij}}, \\
  \label{eq:apb10}
  && b_{ij}= \braket{f}{O_{ij}}.
\end{eqnarray}

\subsection*{Weak formulation of (\ref{eq:c45})--(\ref{eq:c46})}

Using definition (\ref{eq:c38}) of the moments $M_{ij}$ and Eqs. (\ref{eq:apb6})--(\ref{eq:apb7}), the new coefficients $a_{ij}$ and $b_{ij}$ can be expressed in terms of the moments $M_{ij}$ of $f$ through
\begin{equation}
  \label{eq:apb11}
  M_{ij}= \left\{ 
    \begin{array}{lr}
       \sum_{k\le i+j/2}\sum_{l\le i} a_{kl}\braket{e^{-c^2}c^{2k}c_z^l}{E_{kl}} & j\  even \\
       \sum_{k\le i+(j-1)/2}\sum_{l\le i}b_{kl}\braket{e^{-c^2}c^{2k}c_z^l}{O_{kl}} & j\  odd.
    \end{array}
  \right.
\end{equation}
That is, the determination of any function of $\mathcal{M}$ is equivalent to the determination of its moments $M_{ij}$. They can be obtained, and hence the coefficients $a_{ij}$ and $b_{ij}$, imposing Eq. (\ref{eq:c45}) with (\ref{eq:c46}). If Eq. (\ref{eq:c45}) is multiplied by $c^{2i}c_z^j$ and integrated over $\mn{c}$, one has
\begin{eqnarray}
  \label{eq:apb12}
  &&-\beta \left( M_{ij+1}-j M_{02} M_{ij-1}-2iM_{02}M_{i-1j+1}\right)=J_{ij}, \quad i,j\ge 0, \\
  \label{eq:apb13}
&& \phi \ge 0, \qquad M_{10}=\frac{d}{2}, \qquad (M_{00}=1),
\end{eqnarray}
where $J_{ij}$ is defined as
\begin{equation}
  \label{eq:apb14}
  J_{ij}=\int d\mn{c} \ c^{2i}c_z^j J[\mn{c}|\phi].
\end{equation}

\subsection*{Sonine approximation}

Using the first terms of expansion (\ref{eq:apb3}) for $\phi$, we have
\begin{eqnarray}
  \label{eq:apb15}
  \phi(\mn{c})\simeq && \pi^{-d/2} e^{-c^2} \Bigg[a_{00}-a_{01}(c^2-dc_x^2)+ \\
    && \nonumber   +b_{00}c_x+\left(\frac{d-1}{2}b_{01} +\frac{3}{2}
      b_{10}\right) c_x -b_{01}c^2c_x-(b_{10}-b_{01})c_x^3\Bigg].
 \end{eqnarray}
 From a physical point of view, this approximation assumes that the scaled distribution function is an even function of $\mn{c}_\perp=\mn{c}-c_z\mn{u}_z$, it is close to the guassian distribution, and an expansion using Sonine polynomials is properly for the determination of its lowest moments. 

The determination of the coefficients at Eq. (\ref{eq:apb15}) can be achieved using (\ref{eq:apb12}) and (\ref{eq:apb13}). As the system has zero mean velocity ($M_{01}=0$) and $\phi$ is normalized to one ($M_{00}=1$), it is
\begin{equation}
  \label{eq:apb16}
  b_{00}=0,
\end{equation}
and
\begin{equation}
  \label{eq:apb17}
  a_{00}=1.
\end{equation}
Since $M_{10}=\frac{d}{2}$, it is
\begin{equation}
  \label{eq:apb18}
  a_{10}+(d-1)a_{01}=0.
\end{equation}
Finally, for $i=1,j=0$ and $i=0,j=2$, Eq. (\ref{eq:apb12}) leads to
\begin{eqnarray}
  \label{eq:apb19}
  && \beta\frac{1}{4}[3b_{10}+(d-1)b_{01}]=J_{10}, \\
  \label{eq:apb20}
  && \beta\frac{3}{4}b_{10} = J_{02},
\end{eqnarray}
where the collision integrals must be approximately calculated as a function of the coefficients. 

The set of equations (\ref{eq:apb18})--(\ref{eq:apb20}) has more unknown coefficients than equations. In order to circumvent this problem, we can consider the simplest solution denoted with uppercase $^{(0)}$, the one that Chapman-Enskog method takes according to the isotropic properties of the reference state (the local version of the homogeneous cooling state), namely $a_{01}^{(0)}=a_{10}^{(0)}=0$. Consequently, if the collision coefficients, $J_{10}$ and $J_{02}$, are evaluated up to linear orders in $b_{10}$ and $b_{01}$ neglecting contributions from other coefficients, it is
\begin{eqnarray}
  \label{eq:apb21}
  && b_{10}^{(0)}=-\frac{2\sqrt{2}\pi^{(d-1)/2}(1-\alpha^2)}{3d\Gamma\left(\frac{d}{2}\right)\beta}, \\
  \label{eq:apb22}
  && b_{01}^{(0)}=-\frac{2\sqrt{2}\pi^{(d-1)/2}(1-\alpha^2)}{d\Gamma\left(\frac{d}{2}\right)\beta}.
\end{eqnarray}
With this values, we obtain $M_{02}=1/2$ and $M_{11}$ given in Eq. (\ref{eq:c60}).

On the contrary, if we want to evaluate the deviation of $M_{02}$ form $1/2$ we need to compute $a_{01}$. This can be accomplished by adding the following two equations of the hierarchy (\ref{eq:apb12}) to Eqs. (\ref{eq:apb16})--(\ref{eq:apb20}), 
  \begin{equation}
    \label{eq:apb23}
    -\beta (M_{12}-M_{10}M_{02}-2M_{02}^2)= J_{11},
  \end{equation}
and 
  \begin{equation}
    \label{eq:apb24}
    -\beta (M_{04}-3M_{02}^2)= J_{03},
  \end{equation}
and by assuming $a_{20}=a_{11}=a_{02}$. Neglecting nonlinear contribution on the coefficients to the integral collisions, we arrive at 
\begin{eqnarray}
 \label{eq:apb25}
 && a_{01}^{(1)}=-\frac{d(d+2)\Gamma\left(\frac{d}{2}\right)\pi^{-(d-1)/2}}{\sqrt{2}d(1+\alpha)\left[3(1-\alpha)+2d\right]} \beta b_{10},\\
\label{eq:apb26}
  && a_{20}^{(1)}=-\frac{(1+\alpha)\left[16+11d-3\alpha(d+8)\right]}{2\sqrt{2}d(d+2)\Gamma\left(\frac{d}{2}\right)\pi^{-(d-1)/2}}\frac{1}{\beta}b_{10},\\
\label{eq:apb27}
  && b_{01}^{(1)}=b_{10}^{(1)},\\
\label{eq:apb28}
&& b_{10}^{(1)}=-\frac{2\sqrt{2}(1-\alpha^2)d(d+2)\Gamma\left(\frac{d}{2}\right)\pi^{(d-1)/2}\beta^{-1}}{d(d+2)^2\Gamma^2\left(\frac{d}{2}\right)-\frac{3(1-\alpha)(1+\alpha)^2}{16\beta^2}\left[16+11d-3\alpha(d+8)\right]\pi^{d-1}}.
  \end{eqnarray}
Now we use uppercase $^{(1)}$ for the new approximation. If $(1-\alpha)/\beta^2 \ll 1$ and $\alpha \sim 1$, generally verified in simulations, the above relations simplify to 
\begin{eqnarray}
 \label{eq:apb29}
 && a_{01}^{(1)}\simeq\frac{1-\alpha}{d},\\
\label{eq:apb30}
  && a_{20}^{(1)}\simeq\frac{32(d-1)(1-\alpha)\pi^{(d-1)}}{d(d+2)^2\Gamma^2\left(\frac{d}{2}\right)}\frac{1}{\beta^2},\\
\label{eq:apb31}
  && b_{01}^{(1)}\simeq b_{10}^{(1)},\\
\label{eq:apb32}
&& b_{10}^{(1)}\simeq-\frac{4\sqrt{2}(1-\alpha)\pi^{(d-1)/2}}{(d+2)\Gamma\left(\frac{d}{2}\right)}\frac{1}{\beta}.
  \end{eqnarray}
With this latter relations, the contribution to the temperature of the distribution function $M_{02}$ and the reduced heat flux $M_{11}$ reduce to Eqs. (\ref{eq:c59}) and (\ref{eq:c60}), respectively.

\section{Modified event--driven simulations \label{appen:3}}

The usual event-driven algorithms work in two steps: making the system evolve according to the external force in between collisions, and calculating the time collision of events taking place in the system (collisions among particles, collisions among particles and walls, and so on). 

In this work, and since we have a nonuniform external force, the exact computation of the time collisions among particles, for instance, would require the solution of highly nonlinear algebraic equations. This computational problem has been circumvented by simplifying the expression of the force $F(z)$ as follows. As  $F(z)$ varies along the vertical direction, the system was divided into imaginary horizontal layers with height equal a to few values of $\sigma$. If the center of a layer is located at $z=z_0$, then the force is given the constant value $F(z_0)$ throughout the layer. Now, the time calculation of collisions among particles belonging to the same layer needs for the solution of a quadratic algebraic equation, as usual. For particles in two neighbour layers, the computation of time needs for the solution of a quartic algebraic equation. In both cases, the problem is numerically accessible and efficient. As long as the height of the layer leads at the microscopic scale, the approximation of the force is expected to be negligible at meso and macroscales. 

In order to minimize the boundary effects and have a wide bulk, we used thermal boundary conditions for the horizontal walls at the bottom and top of the system. If a particle collides with the wall at the bottom of the system, for instance, it is rebounded with a velocity $\mn{v}=(v_x,v_z)$ given by 
\begin{eqnarray}
  \label{eq:apc1}
&&  v_x=\sqrt{2T_x^+}\sqrt{-\ln (r_1)}, \\
  \label{eq:apc2}
&&  v_z=\sqrt{2T_z^+}\sqrt{-\ln (r_2)}\cos(2\pi r_3),
\end{eqnarray}
where $T_x^+$ and $T_z^+$ are chosen parameters, and $r_1$, $r_2$, $r_3$ are random variables uniformly distributed in $(0,1)$. With this election, the distribution of velocity $\mn{v}$ is borrowed from a gaussian flow crossing the wall at $z=0$ from below, having the two components of $\mn{v}$ different temperatures. Similar expressions were chosen for particles colliding the upper wall (but now with $T_x^-$ and $T_z^-$).

The magnitudes of the simulations are taken following several steps:
\begin{enumerate}
\item The geometry of the system (wight $W$ and height $L$), number of particles ($N$), parameters of the force $(\lambda)$, initial temperature ($T_i$) and boundary conditions ($T_x^{\pm},\ T_z^{\pm}$) are chosen. 
\item The initial state is spatially homogeneous with velocities having gaussian distribution with temperature equal to $T_i$ (usually equal to 1 with respect to unites defined in Sec. \ref{sec4}). 
\item After a transient period, typically after one particle has collide $10^3$ times, the system reaches a the steady state. Measurements are then taken every one collision per particle: number of particles, velocity, energy, $\dots$ at different heights.
\item The magnitudes are averaged over time and over different trajectories or realizations.
\item The density and temperature profiles are analyzed and compared against theoretical predictions for the hydrodynamic profiles (meanly that of Eqs. (\ref{eq:c47}--\ref{eq:c49})). The values of $T_x^{+-}$ and $T_z^{+-}$ are eventually readjusted and we go back to (i). 
\end{enumerate}

\end{document}